\documentstyle[12pt]{article}
\makeatletter
\def\newpic#1{%
   \def\emline##1##2##3##4##5##6{%
      \put(##1,##2){\special{em:point #1##3}}%
      \put(##4,##5){\special{em:point #1##6}}%
      \special{em:line #1##3,#1##6}}}
\newpic{}
\makeatother
\def\d{\partial}

\def\bea{\begin{eqnarray}}
\def\eea{\end{eqnarray}}
\def\nn{\nonumber}
\def\beq{\begin{equation}}
\def\eeq{\end{equation}}
\def\ba{\beq\new\begin{array}{c}}
\def\ea{\end{array}\eeq}
\def\be{\ba}
\def\ee{\ea}
\def\stackreb#1#2{\mathrel{\mathop{#2}\limits_{#1}}}
\def\Tr{{\rm Tr}}
\def\f{1\over}

\makeatletter
\newdimen\normalarrayskip              
\newdimen\minarrayskip                 
\normalarrayskip\baselineskip
\minarrayskip\jot
\newif\ifold             \oldtrue            \def\new{\oldfalse}
\def\arraymode{\ifold\relax\else\displaystyle\fi} 
\def\eqnumphantom{\phantom{(\theequation)}}     
\def\@arrayskip{\ifold\baselineskip\z@\lineskip\z@
     \else
     \baselineskip\minarrayskip\lineskip2\minarrayskip\fi}
\def\@arrayclassz{\ifcase \@lastchclass \@acolampacol \or
\@ampacol \or \or \or \@addamp \or
   \@acolampacol \or \@firstampfalse \@acol \fi
\edef\@preamble{\@preamble
  \ifcase \@chnum
     \hfil$\relax\arraymode\@sharp$\hfil
     \or $\relax\arraymode\@sharp$\hfil
     \or \hfil$\relax\arraymode\@sharp$\fi}}
\def\@array[#1]#2{\setbox\@arstrutbox=\hbox{\vrule
     height\arraystretch \ht\strutbox
     depth\arraystretch \dp\strutbox
     width\z@}\@mkpream{#2}\edef\@preamble{\halign
\noexpand\@halignto
\bgroup \tabskip\z@ \@arstrut \@preamble \tabskip\z@ \cr}%
\let\@startpbox\@@startpbox \let\@endpbox\@@endpbox
  \if #1t\vtop \else \if#1b\vbox \else \vcenter \fi\fi
  \bgroup \let\par\relax
  \let\@sharp##\let\protect\relax
  \@arrayskip\@preamble}
\def\eqnarray{\stepcounter{equation}%
              \let\@currentlabel=\theequation
              \global\@eqnswtrue
              \global\@eqcnt\z@
              \tabskip\@centering
              \let\\=\@eqncr
              $$%
 \halign to \displaywidth\bgroup
    \eqnumphantom\@eqnsel\hskip\@centering
    $\displaystyle \tabskip\z@ {##}$%
    \global\@eqcnt\@ne \hskip 2\arraycolsep
         $\displaystyle\arraymode{##}$\hfil
    \global\@eqcnt\tw@ \hskip 2\arraycolsep
         $\displaystyle\tabskip\z@{##}$\hfil
         \tabskip\@centering
    &{##}\tabskip\z@\cr}
\begingroup\ifx\undefined\newsymbol \else\def\input#1 {\endgroup}\fi
\newfont{\hr}{msbm10}
\newfont{\ams}{msam10}

\font\numbers=cmss12
\font\upright=cmu10 scaled\magstep1
\def\stroke{\vrule height8pt width0.4pt depth-0.1pt}
\def\topfleck{\vrule height8pt width0.5pt depth-5.9pt}
\def\botfleck{\vrule height2pt width0.5pt depth0.1pt}
\def\Zmath{\vcenter{\hbox{\numbers\rlap{\rlap{Z}\kern 0.8pt\topfleck}\kern
2.2pt
                   \rlap Z\kern 6pt\botfleck\kern 1pt}}}
\def\Qmath{\vcenter{\hbox{\upright\rlap{\rlap{Q}\kern
                   3.8pt\stroke}\phantom{Q}}}}
\def\Nmath{\vcenter{\hbox{\upright\rlap{I}\kern 1.7pt N}}}
\def\Cmath{\vcenter{\hbox{\upright\rlap{\rlap{C}\kern
                   3.8pt\stroke}\phantom{C}}}}
\def\Rmath{\vcenter{\hbox{\upright\rlap{I}\kern 1.7pt R}}}
\def\Z{\ifmmode\Zmath\else$\Zmath$\fi}
\def\Q{\ifmmode\Qmath\else$\Qmath$\fi}
\def\N{\ifmmode\Nmath\else$\Nmath$\fi}
\def\C{\ifmmode\Cmath\else$\Cmath$\fi}
\def\R{\ifmmode\Rmath\else$\Rmath$\fi}

\newcounter{app}

\def\app{\setcounter{equation}{0}
\def\theequation{\Alph{app}.\arabic{equation}}\par
   \addvspace{4ex}
   \@afterindentfalse
  \secdef\@app\@dapp}

\newcommand\@app{\@startsection {app}{1}{0ex}%
                                   {-3.5ex \@plus -1ex \@minus -.2ex}%
                                   {2.3ex \@plus.2ex}%
                                   {\normalfont\Large\bf}}
\def\@dapp#1{%
{\parindent \z@ \raggedright  \bf #1}\par\nobreak}
\def\l@app#1#2{\ifnum \c@tocdepth >\z@
    \addpenalty\@secpenalty
    \addvspace{1.0em \@plus\p@}%
    \setlength\@tempdima{8em}%
    \begingroup
      \parindent \z@ \rightskip \@pnumwidth
      \parfillskip -\@pnumwidth
      \leavevmode \bfseries
      \advance\leftskip\@tempdima
      \hskip -\leftskip
      #1\nobreak\hfil \nobreak\hb@xt@\@pnumwidth{\hss #2}\par
    \endgroup\fi}
\newcounter{sapp}[app]

\def\sapp{\def\theequation{\Alph{app}.\arabic{equation}}
\par
\@afterindentfalse
  \secdef\@sapp\@dsapp}
\newcommand{\@sapp}{\@startsection{sapp}{2}{\z@}%
                                     {-3.25ex\@plus -1ex \@minus -.2ex}%
                                     {1.5ex \@plus .2ex}%
                                     {\normalfont\large\bfseries}}

\def\@dsapp#1{%
{\parindent \z@ \raggedright  \bf #1
}\par\nobreak}
\newcommand{\l@sapp}{\@dottedtocline{2}{1.5em}{2.3em}}

\def\stackreb#1#2{\mathrel{\mathop{#2}\limits_{#1}}}
\def\Tr{{\rm Tr}}

\def\d{\partial}

\def\f{1\over}

\def\2{{1\over 2}}
\def\N2{${\cal N}=2$}

\def\be{ \begin{eqnarray} }
\def\ee{ \end{eqnarray} }

\def\d{\partial}

\def\bea{\begin{eqnarray}}
\def\eea{\end{eqnarray}}
\def\nn{\nonumber}

\def\beq{\begin{equation}}
\def\eeq{\end{equation}}
\def\ba{\beq\new\begin{array}{c}}
\def\ea{\end{array}\eeq}
\def\be{\ba}
\def\ee{\ea}
\def\stackreb#1#2{\mathrel{\mathop{#2}\limits_{#1}}}
\def\f{1\over}

\def\P{\bf P}

\begin{document}
\begin{flushright}
ITEP/TH-11/99\\
FIAN/TD-01/99\\
hepth/9903088
\end{flushright}
\vspace{0.5cm}
\begin{center}
{\LARGE \bf WDVV Equations and Seiberg-Witten theory}\hspace{0.05cm}
\footnote{Talk delivered
at the Workshop "Integrability: the Seiberg-Witten and Whitham equations",
Edinburgh, September 1998.}
\vspace{0.5cm}

\setcounter{footnote}{1}
\def\thefootnote{\fnsymbol{footnote}}
{\Large A.Mironov\footnote{Theory
Department, Lebedev Physics Institute, Moscow
~117924, Russia; e-mail address: mironov@lpi.ac.ru}\footnote{ITEP,
Moscow 117259, Russia; e-mail address:
mironov@itep.ru}
}\\
\end{center}
\bigskip
\begin{quotation}
We present a review of the results on the associativity
algebras and WDVV equations associated with the Seiberg-Witten solutions of
$N=2$ SUSY gauge theories. It is mostly based on the
integrable treatment of these solutions.
We consider various examples of the
Seiberg-Witten solutions and corresponding integrable systems
and discuss when the WDVV equations hold. We also discuss a covariance of
the general WDVV equations.
\end{quotation}
\setcounter{footnote}{0}
\setcounter{equation}{0}

\section{What is WDVV}
More than two years ago N.Seiberg and E.Witten \cite{SW} proposed a new way
to deal with the low-energy effective actions of $N=2$ four-dimensional
supersymmetric gauge theories, both pure gauge theories (i.e. containing
only vector supermultiplet) and those with matter hypermultiplets. Among
other things, they have shown that the low-energy effective actions (the
end-points of the renormalization group flows) fit into universality classes
depending on the vacuum of the theory.  If the moduli space of these vacua is
a finite-\-dimensional variety, the effective actions can be essentially
described in terms of a system with {\it finite}-dimensional phase space (\# of
degrees of freedom is equal to the rank of the gauge group), although the
original theory lives in a many-dimen\-si\-on\-al space-time.  These effective
theories turn out to be integrable. Integrable structures behind the
Seiberg-Witten (SW) appro\-ach has been found in \cite{GKMMM} and later
examined in detail for different theories in \cite{MW}-\cite{rev}.

The second important property of the SW fra\-me\-work
which merits the adjective
"topological" has been
more recently revealed in the series of papers
\cite{WDVVa}-\cite{BMMM1}
and has
much to do with the associative algebras. Namely, it turns out that the
prepotential of SW theory satisfies a set of
Witten-Dijkgraaf-Verlinde-Verlinde (WDVV) equations. These equations have
been originally presented in \cite{WDVV} (in a different form, see below)
\be\label{wdvv}
F_iF_j^{-1}F_k=F_kF_j^{-1}F_i
\ee
where $F_i$'s are matrices with the matrix elements that are the third
derivatives of the unique function $F$ of many variables $a_i$'s
(prepotential in the SW theory) parameterizing a moduli space:
\be
\left(F_i\right)_{jk}\equiv {\partial^3 F\over \d a_i\d a_j\d a_k},
\ \ \ \ i,j,k=1,...,n
\ee
Although generally there is a lot of solutions to the matrix equations
(\ref{wdvv}), it is extremely non-trivial task to express all the matrix
elements through the only function $F$. In fact, there have been only known
the two different classes of the non-trivial solutions to the WDVV equations,
both being intimately related to the two-dimensional topological theories of
type A (quantum cohomologies \cite{typeA}) and of type B ($N=2$ SUSY
Landau-Ginzburg (LG) theories that were investigated in a variety of papers,
see, for example, \cite{typeB} and references therein). Thus, the existence
of a new class of solutions connected with the four-dimensional theories
looks quite striking.  It is worth noting that both the two-dimensional
topological theories and the SW theories reveal the integrability structures
related to the WDVV equations.  Namely, the function $F$ plays the role of
the (quasiclassical) $\tau$-function of some Whitham type hierarchy
\cite{typeB,GKMMM,whit}.

In this brief review, we will describe the results of papers
\cite{WDVVa}-\cite{BMMM1}
that deal
with the structure and origin of the WDVV equations in the SW theories and,
to some extent, with their general properties\footnote{We tried to make this
review self-consistent. Some related points can be found in other talks
presented at the Workshop, in particular, delivered by A.Marshakov and
A.Morozov.}.  To give some insight of the general structure of the WDVV
equations, let us consider the simplest non-trivial examples of $n=3$ WDVV
equations in topological theories. The first example is the $N=2$ SUSY LG
theory with the superpotential $W'(\lambda)=\lambda^3-q$ \cite{typeB}.  In
this case, the prepotential reads as
\be
F=\2 a_1a_2^2+\2 a_1^2a_3+{q\over
2}a_2a_3^2
\ee
and the matrices $F_i$ (the third derivatives of the
prepotential) are
\be
F_1 = \left(\begin{array}{ccc} 0&0&1\\0&1&0\\1&0&0
\end{array}\right), \ \ \ \ F_2 = \left(\begin{array}{ccc}
0&1&0\\1&0&0\\0&0&q\end{array}\right),
\ \ \ \
F_3 = \left(\begin{array}{ccc}
1&0&0\\0&0&q\\0&q&0 \end{array}\right).
\label{Fexpl}
\ee
One can easily check that these matrices do really satisfy the WDVV equations
(\ref{wdvv}).

The second example is the quantum cohomologies of C$\P ^2$.
In this case, the prepotential is given by the formula \cite{typeA}
\be
F=\2 a_1a_2^2+\2 a_1^2a_3+\sum_{k=1}^{\infty}{N_ka_3^{3k-1}\over
(3k-1)!} e^{ka_2}
\ee
where the coefficients $N_k$ (describing the rational Gromov-Witten classes)
counts the number of the rational curves in
C$\P ^2$ and are to be calculated. Since the matrices $F$ have the form
\be
F_1 = \left(\begin{array}{ccc}
0&0&1\\0&1&0\\1&0&0\end{array}\right), \ \ \ \
F_2 = \left(\begin{array}{ccc}
0&1&0\\1&F_{222}&F_{223}\\0&F_{223}&F_{233}\end{array}\right),
\ \ \ \
F_3 = \left(\begin{array}{ccc}
1&0&0\\0&F_{223}&F_{233}\\0&F_{233}&F_{333}\end{array}\right)
\ee
the WDVV equations are equivalent to the identity
\be
F_{333}=F_{223}^2-F_{222}F_{233}
\ee
which, in turn, results into the recurrent relation defining the coefficients
$N_k$:
\be
\frac{N_k}{(3k-4)!} = \sum_{a+b=k}
\frac{a^2b(3b-1)b(2a-b)}{(3a-1)!(3b-1)!}N_aN_b.
\ee
The crucial feature of the presented examples is that, in both cases, there
exists a constant matrix $F_1$. Following \cite{typeB},
one can consider it as a flat
metric on the moduli space. In fact, in its original version, the WDVV
equations have been written in a slightly different form, that is, as
the associativity condition of some algebra. We will discuss this later, and
now just remark that, having distinguished (constant) metric $\eta\equiv F_1$,
one can naturally rewrite (\ref{wdvv}) as the equations
\be\label{Cas}
C_iC_j=C_jC_i
\ee
for the matrices $C_i\equiv
\eta^{-1} F_i$, i.e. $\left(C_i\right)^j_k=\eta^{jl}F_{ilk}$. Formula
(\ref{Cas}) is equivalent to (\ref{wdvv}) with $j=1$. Moreover, this
particular relation is already sufficient \cite{WDVVlong,WDVVr} to reproduce
the whole set of the WDVV equations (\ref{wdvv}). Indeed, since $F_i=F_1
C_i$, we obtain
\be
F_iF^{-1}_jF_k=F_1\left(C_iC^{-1}_kC_j\right)
\ee
which is obviously symmetric under the permutation $i\leftrightarrow j$.
Let us also note that, although the WDVV equations can be fulfilled only for
some specific choices of the coordinates $a_i$ on the moduli space, they
still admit any linear transformation. This defines the flat structures on
the moduli space, and we often call $a_i$ flat coordinates.

In fact, the existence of the flat metric is not necessary for
(\ref{wdvv}) to be true, how we explain below. Moreover, the SW theories give
exactly an example of such a case, where there is no distinguished constant
matrix. This matrix can be found in topological theories because of
existence their field theory interpretation where the unity operator is always
presented.

\section{Perturbative SW prepotentials}
Before going into the discussion of the WDVV equations for the complete SW
prepotentials, let us note that the leading perturbative part of them should
satisfy the equations (\ref{wdvv}) by itself (since the classical
quadratic piece does not contribute into the third derivatives). In each
case it can be checked by the straightforward calculation. On the other hand,
if the WDVV equations are fulfilled for perturbative prepotential, it is
a necessary condition for them to hold for complete prepotential.

The perturbative prepotential can be obtained from the
one-loop field theory
calculations. To this end, let us note that
there are two origins of masses in $N=2$ SUSY YM (SYM) models:
first, they can be generated by vacuum values of the scalar $\phi$ from  the
gauge supermultiplet. For a supermultiplet in representation $R$
of the gauge group $G$ this contribution to the prepotential
is given by the analog of the Coleman-Weinberg formula
(from now on, we omit the classical part of the prepotential from all
expressions):
\be
F_R = \pm\frac{1}{4} \Tr_R \ \phi^2\log\phi,
\label{ColeWei}
\ee
and the sign is ``$+$'' for vector supermultiplets (normally they
are in the adjoint representation) and ``$-$'' for matter hypermultiplets.
Second, there are bare masses $m_R$ which should be added to $\phi$
in (\ref{ColeWei}). As a result, the general expression for the
perturbative prepotential is
\be
F = \frac{1}{4}\sum_{{vector}\atop{mplets}} \Tr_{A}
(\phi + M_nI_A)^2\log(\phi + M_nI_A) -
\\
- \frac{1}{4}\sum_{{hyper}\atop{mplets}} \Tr_R
(\phi + m_RI_R)^2\log(\phi + m_RI_R) + f(m)
\label{PertuF}
\ee
where the term $f(m)$ depending only on masses is not fixed by the
(perturbative) field theory but can be read off from the
non-perturbative description, and $I_R$ denotes the unit matrix in the
representation $R$.

As a concrete example, let us consider the $SU(n)$ gauge group. Then, say,
perturbative prepotential for the pure gauge theory acquires the
form
\be
F_V^{pert}={\f 4}\sum_{ij}
\left(a_i-a_j\right)^2\log\left(a_i-a_j\right)
\ee
This formula establishes that when v.e.v.'s
of the scalar fields in the gauge supermultiplet are non-vanishing
(perturbatively $a_r$ are eigenvalues of the vacuum
expectation matrix  $\langle\phi\rangle$), the fields in the gauge multiplet
acquire masses $m_{rr'} = a_r - a_{r'}$ (the pair of indices $(r,r')$ label
a field in the adjoint representation of $G$). In the $SU(n)$ case,
the eigenvalues are subject to the condition $\sum_ia_i=0$.
Analogous formula for the
adjoint matter contribution to the prepotential is
\be
F_A^{pert}=-{\f 4}\sum_{ij}
\left(a_i-a_j+M\right)^2\log\left(a_i-a_j+M\right)
\ee
while the contribution of the fundamental matter reads as
\be
F_F^{pert}=-{\f 4}\sum_{i}
\left(a_i+m\right)^2\log\left(a_i+m\right)
\ee

Similar formulas can be obtained for the other groups.
The eigenvalues of $\langle\phi\rangle$
in the first
fundamental representation of the classical series of the Lie groups are
\be
B_n\ (SO(2n+1)):\ \ \ \ \ \{a_1,...,a_n,0,-a_1,...,-a_n\};\\
C_n\ (Sp(n)):\ \ \ \ \ \{a_1,...,a_n,-a_1,...,-a_n\};\\
D_n\ (SO(2n)):\ \ \ \ \ \{a_1,...,a_n,-a_1,...,-a_n\}
\ee
while the eigenvalues in the adjoint representation have the form
\be\label{adj}
B_n:\ \ \ \ \ \{\pm a_j;\pm a_j\pm a_k\};\ \ \ j<k\le n\\
C_n:\ \ \ \ \ \{\pm 2a_j;\pm a_j\pm a_k\};\ \ \ j<k\le n\\
D_n:\ \ \ \ \ \{\pm a_j\pm a_k\}, \ \ j<k\le n
\ee
Analogous formulas can be written for the exceptional groups too.  The prepotential in the pure gauge theory can be
read off from the formula (\ref{adj}) and has the form
\be\label{adjvo}
B_n:\ \ \ \ \ F_0={1\over 4}\sum_{i,i} \left(\left(a_{i}-a_{j}
\right)^2\log\left(a_{i}-a_{j}\right)+\left(a_{i}+a_{j}
\right)^2\log\left(a_{i}+a_{j}\right)\right)+
{1\over 2}\sum_i a_i^2\log a_i;\\
C_n:\ \ \ \ \ F_0={1\over 4}\sum_{i,i} \left(\left(a_{i}-a_{j}
\right)^2\log\left(a_{i}-a_{j}\right)+\left(a_{i}+a_{j}
\right)^2\log\left(a_{i}+a_{j}\right)\right)
+2\sum_i a_i^2\log a_i;\\
D_n:\ \ \ \ \ F_0={1\over 4}\sum_{i,i} \left(\left(a_{i}-a_{j}
\right)^2\log\left(a_{i}-a_{j}\right)+\left(a_{i}+a_{j}
\right)^2\log\left(a_{i}+a_{j}\right)\right)
\ee

The perturbative prepotentials are discussed in detail in
\cite{WDVVlong}. In that paper is also contained the proof of the WDVV
equations for these prepotentials. Here we just list some results of
\cite{WDVVlong} adding some new statements recently obtained.

{\bf i)} The WDVV equations always hold for the pure gauge theories
$F^{pert}=F_V^{pert}$ (including the exceptional groups)
\footnote{The rank of
the group should be bigger than 2 for the WDVV equations not to be empty,
thus for example in the pure gauge $G_2$-model they are satisfied trivially.
It have been checked in \cite{WDVVlong,IY} (using MAPLE package)
that the WDVV
equations hold for the perturbative prepotentials of the pure gauge $F_4$,
$E_6$ and $E_7$ models.
Note that the corresponding non-perturbative SW curves are not
obviously hyperelliptic.}. In fact, in \cite{WDVVlong} it has been proved
that, if one starts with the general prepotential of the form
\be
F={1\over 4}\sum_{i,i} \left(\alpha_-\left(a_{i}-a_{j}
\right)^2\log\left(a_{i}-a_{j}\right)+\alpha_+\left(a_{i}+a_{j}
\right)^2\log\left(a_{i}+a_{j}\right)\right)+{\eta\over 2}\sum_i a_i^2\log a_i
\ee
the WDVV hold iff $\alpha_+=\alpha_-$ or $\alpha_+=0$, $\eta$ being
arbitrary. A new result due to A.Veselov \cite{Ves} shows that this form can
be further generalized by adding boundary terms. For this latter case,
however, we do not know the non-perturbative extension.

{\bf ii)} If one considers the gauge supermultiplets interacting with the
$n_f$
matter hypermultiplets in the first fundamental representation with masses
$m_{\alpha}$
\be\label{complete}
F^{pert}=F_V^{pert}+rF_F^{pert}+Kf_F(m)
\ee
(where $r$ and $K$
are some undetermined coefficients), the WDVV equations do not hold unless
$K=r^2/4$, the masses are regarded
as moduli (i.e. the equations (\ref{wdvv}) contain the derivatives with
respect to masses) and
\be
f_F(m)= {\f
4}
\sum_{\alpha,\beta}\left(\left(m_{\alpha}-m_{\beta}
\right)^2\log\left(m_{\alpha}-m_{\beta}\right)
\right)
\ee
for the $SU(n)$ gauge group and
\be
f_F(m)= {\f
4}
\sum_{\alpha,\beta}\left(\left(m_{\alpha}-m_{\beta}
\right)^2\log\left(m_{\alpha}-m_{\beta}\right)+\left(m_{\alpha}+m_{\beta}
\right)^2\log\left(m_{\alpha}+m_{\beta}\right)\right)+\\+
{r(r+s)\over 4}
\sum_{\alpha}m_{\alpha}^2\log m_{\alpha}
\ee
for other classical groups, $s=2$ for the orthogonal groups and
$s=-2$ for the symplectic ones.

Note that at value $r=-2$ the prepotential (\ref{complete}) can be considered
as that in the pure gauge theory with the gauge group of the higher rank
$\hbox{rank}G+n_f$. At the same time,
at value $r=2$, like $a_i$'s lying in irrep of $G$, masses
$m_{\alpha}$'s can be regarded as lying in irrep of some $\widetilde G$ so
that if $G=A_n$, $C_n$, $D_n$, $\ \widetilde G=A_n$, $D_n$, $C_n$
accordingly (this is nothing but the notorious (gauge group
$\longleftrightarrow$ flavor group)
duality, see, e.g. \cite{dual}). These correspondences "explain" the form of
the mass term in the prepotential $f(m)$.

{\bf iii)} The set of the perturbative prepotentials satisfying the WDVV
equations can be further extended. Namely, one can consider higher
dimensional SUSY gauge theories
\cite{N,hd,BMMM2,hd1}, in particular, $5d$ theories
compactified onto the circle of radius $R$, so that in
four-dimensions it can be seen as a gauge theory of the infinitely many
vector supermultiplets with masses $M_k=\pi k/R$.
Then, the perturbative prepotentials in the pure gauge $SU(n)$ theory
of such a type reads as
\be\label{FARTC}
F^{pert}={1\over 4}\sum_{i,j}\left({i
\over 3}a_{ij}^3+{1\over 2}
\hbox{Li}_3\left(e^{-2iRa_{ij}}\right)\right)-
{in\over 6}\sum_{i>j>k}a_i^3
\ee
where $a_{ij}\equiv a_i-a_j$ and
$\hbox{Li}_3(x)$ is the standard three-loga\-rithm
function. The first sum in this expression tends to the usual logarithmic
prepotential $F_V^{pert}$ as $R\to 0$, while the second one vanishes. It
deserves mentioning that the second (cubic) term do not come from any field
theory calculation, but corresponds to the Chern-Simons term
$\Tr\left(A\wedge F\wedge F\right)$ in the field theory Lagrangian
\cite{Sei}. It is similar to the $U^3$-terms of the perturbative
prepotential $F^{pert}$ of the heterotic string \cite{HM}. The presence of
these terms turns to be absolutely crucial for the WDVV equations to hold.
Further details on this case, and on the prepotential with
fundamental hypermultiplets included can be found in \cite{hd1}.

One can also consider other classical groups. Then, the perturbative
prepotentials acquire the form (\ref{adjvo}) with all $x^2\log x$ substituted
by ${i\over 3}x^3+{1\over 2} \hbox{Li}_3\left(e^{-2iRx}\right)$. One can
easily check, along the line of \cite{WDVVlong} that these prepotentials
satisfy the WDVV equations.

{\bf iv)} If in the $4d$ theory the adjoint matter hypermultiplets are
presented, i.e. $F^{pert}=F_V^{pert}+F_A^{pert}+f_A(m)$, the WDVV equations
never hold. At the same time, the WDVV equations are fulfilled for the
theory with matter hypermultiplets in the symmetric/antisymmetric
square of the fundamental representation\footnote{These hypermultiplets
contribute to the prepotential
\be
F_{S} =  -\frac{1}{4}\sum_{i\leq j} (a_i+a_{j}+m)^2
\log(a_i+a_{j} +m)\\
F_{AS} =  -\frac{1}{4}\sum_{i < j} (a_i+a_{j}+m)^2
\log(a_i+a_{j} +m)
\ee
} iff the masses of these hypermultiplets are equal to zero.

{\bf v)} Our last example \cite{BMMM1,BMMM2}
has the most unclear status, at the moment. It
corresponds to the pure gauge $5d$ theory with higher, $N=2$ SUSY in
five dimensions. Starting
with such a five dimensional model
one may obtain four dimensional $N=2$ SUSY models (with fields only
in the adjoint representation of the gauge group)
by imposing non-trivial boundary conditions on half of the fields:
\be
\phi(x_5 +R) = e^{2i\epsilon}\phi(x_5).
\label{bc}
\ee
If $\epsilon = 0$ one obtains $N=4$ SUSY in four dimensions,
but when $\epsilon \neq 2\pi n$ this is explicitly broken to $N=2$.
The low-energy mass spectrum of the  four dimensional theory this time contains
two towers of Kaluza-Klein modes:
\be
M = \frac {\pi n}{R}\ \ {\rm and} \ \
M = \frac{\epsilon + \pi n}{R}, \ \ \ N\in { Z}.
\label{spectrum}
\ee
The prepotential for the group $SU(n)$ ($i = 1\ldots n$) should be
\be\label{pp}
{F}^{pert}={1\over 4}\
\sum_{i,j}\hspace{-0.2cm}{\phantom{a}}^{'} \sum_{n=-\infty}^\infty
\Bigg\lbrace \left(Ra_{ij}+\pi n \right)^2 \log\left(Ra_{ij}+\pi n\right)\\
\qquad -\left(Ra_{ij}+\pi n-\epsilon\right)^2
\log\left(Ra_{ij}+\pi n-\epsilon\right)\Bigg\rbrace
={1\over 4}\sum_{i,j}\hspace{-0.2cm}{\phantom{a}}^{'}f(a_{ij})
\ee
with
\be
f(a)=\hbox{Li}_3\left(e^{-2iRa}\right)
-\hbox{Li}_3\left(e^{-2i(Ra+\epsilon)}\right)
\ee

The prepotential (\ref{pp}) satisfies the WDVV equations iff $\epsilon=\pi$
\cite{BMMM1}. Moreover, it gives the most general solution
for a general class of perturbative prepotentials
${F}_{pert}$ assuming the functional form
\be\label{funform}
{F}=\sum_{\alpha\in \Phi}f(\alpha\cdot a),
\ee
where the sum is over the root system $\Phi$ of a Lie algebra.
The mystery about this prepotential is that, on one hand, it never satisfies
the WDVV equations unless $\epsilon=\pi$ and we do not know if the
corresponding complete non-perturbative prepotential satisfies the WDVV even
for $\epsilon=\pi$. On the other hand, in the limit $R \rightarrow 0$ and
$\epsilon \sim mR$ for finite $m$, (when the mass spectrum (\ref{spectrum})
reduces to the two points $M = 0$ and $M = m$), the theory is the four
dimensional YM model with $N=4$ SUSY softly broken to $N=2$, i.e. includes
the adjoint matter hypermultiplet. It corresponds to item {\bf iv)} when the
WDVV always do {\it not} hold. By all these reasons, this exceptional case
deserves further investigation.

From the above consideration of the WDVV equations for the perturbative
prepotentials, one can learn the following lessons:
\begin{itemize}
\item
masses are to be regarded as moduli
\item
as an empiric rule, one may say that the WDVV
equations are satisfied by perturbative prepotentials which depend only on
the pairwise sums of the type $(a_i\pm b_j)$, where moduli $a_i$ and $b_j$ are
either periods or masses\footnote{This general rule can be
easily interpreted in D-brane terms, since the interaction of branes
is caused by strings between them. The pairwise structure $(a_i\pm
b_j)$ exactly reflects this fact, $a_i$ and $b_j$ should be identified with
the ends of string.}. This is the case for the models that contain either
massive matter hypermultiplets in
the first fundamental representation (or its dual), or massless
matter in the square product of those.
Troubles arise in all other situations because of the terms
with $a_i\pm b_j\pm c_k\pm\ldots$. (The inverse statement is wrong --
there are some exceptions when the WDVV equations hold despite the presence
of such terms -- e.g., for the exceptional groups.)
\end{itemize}

Note that, for the non-UV-finite theories with the perturbative prepotential
satisfying the WDVV equations, one can add one more parameter to the set of
moduli -- the parameter $\Lambda$ that enters all the logarithmic terms as
$x^2\log x/\Lambda$. Then, some properly defined WDVV equations still
remain correct despite the matrices $F_i^{-1}$ no longer exist (one just
needs to consider instead of them the matrices of the proper minors)
\cite{BM} -- see footnote 9 in sect.6.

\section{Associativity conditions}
In the context of the two-dimensional LG topological theories, the
WDVV equations arose as associativity condition of some polynomial algebra.
We will prove below that the equations in the SW theories have the same
origin. Now we briefly remind the main ingredients of this approach in the
standard case of the LG theories.

In this case, one deals with the chiral ring formed by a set of polynomials
$\left\{\Phi_i(\lambda)\right\}$ and two co-prime (i.e. without common
zeroes) fixed polynomials $Q(\lambda)$ and $P(\lambda)$. The polynomials
$\Phi$ form the associative algebra with the structure constants $C_{ij}^k$
given with respect to the product defined by modulo $P'$:
\be
\Phi_i\Phi_j=C_{ij}^k\Phi_kQ'+(\ast)P'\longrightarrow C_{ij}^k\Phi_kQ'
\ee
the associativity condition being
\be
\left(\Phi_i\Phi_j\right)\Phi_k=\Phi_i\left(\Phi_j\Phi_k\right),
\ee
\be\label{ass}
\hbox{ i.e. }\ \ \
C_iC_j=C_jC_i,\ \ \ \left(C_i\right)_k^j=C_{ik}^j
\ee
Now, in order to get from these conditions the WDVV equations, one needs to
choose properly the flat moduli \cite{typeB}:
\be
a_i=-{n\over i(n-i)}\hbox{res}\left(P^{i/n}dQ\right),\ \ \ n=\hbox{ord} (P)
\ee
Then, there exists the prepotential whose third derivatives are given by
the residue formula
\be\label{res}
F_{ijk}=
\stackreb{P' = 0}{\hbox{res}}
\frac{\Phi_i\Phi_j\Phi_k}{P'}
\ee
On the other hand, from the associativity condition (\ref{ass}) and
the residue formula (\ref{res}), one obtains that
\be\label{im}
F_{ijk}=\left(C_i\right)_j^lF_{Q'lk},\ \ \hbox{ i.e. }\ \ C_i=F_iF^{-1}_{Q'}
\ee
Substituting this formula for $C_i$ into (\ref{ass}), one finally reaches the
WDVV equations in the form
\be
F_i G^{-1} F_j = F_j G^{-1} F_i,\\
G \equiv F_{Q'}
\label{WDVVgen}
\ee
The choice $Q'=\Phi_l$ gives the standard
equations (\ref{wdvv}). In two-dimensional topological theories, there is
always the unity operator that corresponds to $Q'=1$ and leads to the
constant metric $F_{Q'}$.

Thus, from this short study of the WDVV equations in the LG theories, we can
get three main ingredients necessary for these equations to hold. These are:
\begin{itemize}
\item
associative algebra
\item
flat moduli (coordinates)
\item
residue formula
\end{itemize}
We will show that in the SW theories only the first ingredient requires
a non-trivial check, while the other two are automatically presented due to
proper integrable structures.

\section{SW theories and integrable systems}
Now we turn to the WDVV equations emerging within the context of
the SW construction \cite{SW}
and show how
they are related to integrable system underlying the corresponding SW theory.
The most
important result of
\cite{SW}, from this point of view, is that the moduli space of
vacua and low energy effective action in SYM theories
are completely given by the following input data:
\begin{itemize}
\item
Riemann surface ${\cal C}$
\item
moduli space ${\cal M}$ (of the curves ${\cal C}$)
\item
meromorphic 1-form $dS$ on ${\cal C}$
\end{itemize}
How it was pointed out in \cite{GKMMM,WDVVlong,GMMM},
this input can be naturally described
in the framework of some underlying integrable system. Let us consider a
concrete example -- the $SU(n)$ pure gauge SYM theory that
is associated with the periodic Toda chain with $n$ sites.
This integrable system is
entirely given by the Lax operator
\be\label{Lax}
L(w)=\left(\begin{array}{cccc}
p_1 & e^{q_1-q_2} & & w\\
e^{q_1-q_2} & p_2 & \vdots&\\
&\ldots&\ddots &\vdots\\
{\f w}&& \ldots&p_n
\end{array}\right)
\ee
The Riemann surface ${\cal C}$ of the SW data is nothing but the spectral
curve of the integrable system, which is given by the equation
\be
\det\left(L(w)-\lambda\right)=0
\ee
Taking into account (\ref{Lax}), one can get from this formula the
equation
\begin{equation}\label{scurve}
w+{\f w}=P\left(\lambda\right)=\prod_{i=1}^n\left(\lambda-\lambda_i\right),
\ \ \ \sum_i\lambda_i=0
\end{equation}
where the ramification points $\lambda_i$ are Hamiltonians (integrals
of motion) parameterizing the moduli space ${\cal M}$ of the spectral curves.
The replace $Y\equiv w-1/w$ transforms the curve (\ref{scurve}) to the
standard hyperelliptic form $\ Y^2=P^2-4$, the genus of the curve
being $n-1$.                                 

The same integrable system, i.e. the periodic Toda chain can be
alternatively
rewritten in terms of
the $2\times 2$ Lax matrices ${\cal L}_i$ each associated with the site
of the chain:
\be\label{TodaLax}
{\cal L}_i=\left(
\begin{array}{cc}
\lambda+p_i&e^{q_i}\\
-e^{-q_i}&0
\end{array}
\right)
\ee
The Lax operator ${\cal L}_i$ can be considered as an "infinitesimal"
transfer matrix that shifts from the $i$-th to the $i+1$-th site of the
chain
\be\label{lproblem}
{\cal L}_i(\lambda)\Psi_i(\lambda)=\Psi_{i+1}(\lambda)
\ee
where $\Psi_i(\lambda)$ is the two-component Baker-Akhiezer function.

One also needs to consider proper
boundary conditions. In the $SU(n)$
case, they are periodic.
The periodic boundary conditions are easily formulated in terms
of the Baker-Akhiezer function and read as
\be\label{pbc}
\Psi_{i+N_c}(\lambda)=w\Psi_{i}(\lambda)
\ee
where $w$ is a free parameter (diagonal matrix).
The Toda chain with these boundary conditions can be naturally associated with
the Dynkin diagram of the group $A_{n-1}^{(1)}$.

One can also introduce the transfer matrix shifting $i$ to $i+n$
\be\label{Tmat}
T(\lambda)\equiv {\cal L}_{n}(\lambda)\ldots {\cal L}_1(\lambda)
\ee
Now the periodic boundary conditions are encapsulated in
the spectral curve equation
\be\label{specurv}
\det (T(\lambda)-w\cdot {\bf 1})=0
\ee
or
\be\label{Todasc}
w^2-\Tr T(\lambda) w+\det T(\lambda)=0
\ee
This curve coincides with (\ref{scurve}), since $\det T(\lambda)=1$ for the
Toda Lax operators (\ref{TodaLax}).

The last important ingredient of the construction is the meromorphic 1-form
$dS=\lambda{dw\over w}=\lambda {dP\over Y}$. From the point of view of
the Toda chain, it is just the shortened action "$pdq$" along the
non-contractible contours on the Hamiltonian tori. Its defining property is
that the derivatives of $dS$ with respect to the moduli (ramification points)
are holomorphic differentials on the spectral curve.                        

After this concrete example, we are ready to
describe how the SW data emerge within
a more general integrable framework and then discuss more on the
concrete examples of the SW construction.
As before, we start with the theories without matter
hypermultiplets. First,                                                                                                                                                                                                                                                                                                                                                                                                                                                                      matter hypermultiplets. First,
we introduce bare spectral curve $E$ that is torus
$y^2=x^3+g_2x^2+g_3$ for the UV finite
SYM theories with the associated holomorphic 1-form
$d\omega=dx/y$. This bare spectral curve degenerates into the
double-punctured sphere (annulus) for the asymptotically free theories: $x\to
w+1/w$, $y\to w-1/w$, $d\omega=dw/w$.
On this bare curve, there are given either a
matrix-valued Lax operator $L(x,y)$ if one considers an extension of the
(\ref{Lax}) Lax representation, or another matrix Lax operator ${\cal
L}_i(x,y)$ associated with an extension of the representation (\ref{TodaLax})
and defining the transfer matrix $T(x,y)$. The corresponding dressed spectral
curve is defined either from the formula $\det(L-\lambda)=0$, or from
$\det(T-w)=0$.

This spectral curve is a
ramified covering of $E$ given by the equation
\be
{\cal P}(\lambda;x,y)=0
\ee
In the case of the gauge group  $G=SU(n)$, the function ${\cal P}$ is a
polynomial of degree $n$ in $\lambda$.

Thus, the moduli space ${\cal M}$ of the spectral curve is given just  by
coefficients of ${\cal P}$.
The generating 1-form $dS \cong \lambda d\omega$ is meromorphic on
${\cal C}$ ("$\cong$" denotes the equality modulo total derivatives).

The prepotential and other "physical" quantities are defined in terms of the
cohomology class of $dS$:
\be
a_i = \oint_{A_i} dS,\\ a_i^D\equiv {\d F\over\d a_i}=\oint_{B_i}dS,\\
\ A_I \circ B_J = \delta_{IJ}.
\label{defprep}
\ee
The first identity defines here the appropriate flat moduli, while the second
one -- the prepotential. The defining property
of the generating differential
$dS$ is that its derivatives w.r.t. moduli
give holomorphic 1-differentials. In particular,
\be
{\d dS\over \d a_i}=d\omega_i
\ee
and, therefore, the second derivative of the prepotential
w.r.t. $a_i$'s is the period
matrix of the curve ${\cal C}$:
\be
{\d^2F\over\d a_i\d a_j}=T_{ij}
\ee
The latter formula allows one to identify prepotential with logarithm of the
$\tau$-function of the Whitham hierarchy \cite{typeB,whit}: $F=\log\tau$.

So far we reckoned without matter hypermultiplets.
In order to include them, one just needs to consider the
surface ${\cal C}$ with punctures. Then, the masses are proportional to
residues of $dS$ at the punctures, and the moduli space has to be extended to
include these mass moduli. All other formulas remain in essence the same
(see \cite{WDVVlong} for more details).

By the present moment, the correspondence between SYM theories and integrable
systems is built through the SW construction in most of known cases that
are collected in the table\footnote{In the table we considered only the
classical groups.}.

\begin{figure}[t]
\begin{center}
{\large {\bf Table.}
SUSY gauge theories $\Longleftrightarrow$ integrable systems
correspondence}
\end{center}
\begin{center} \begin{tabular}{|c|c|c|c|c|}
\hline
SUSY    & Pure gauge      & SYM theory & SYM theory     \\
physical& SYM theory,     & with fund. & with adj.      \\
theory  & gauge group $G$ & matter     & matter         \\
\hline
        & Toda chain      & $XXX$      & Calogero-Moser \\
 $4d$   & for the dual    & spin       &     system     \\
        & affine ${\hat G}^{\vee}$ & chain &            \\
\hline
        & Relativistic    & $XXZ$      & Ruijsenaars-   \\
 $5d$   & Toda            & spin       & Schneider      \\
        & chain           & chain      & model\footnote{This case is better
associated with the specific boundary conditions imposed on the fields in the
fifth dimension (see \cite{BMMM1,BMMM2} and sect.2) than with the adjoint
matter added \cite{N}.}          \\
\hline
        &                 & $XYZ$      &                \\
 $6d$   &                 & spin       &                \\
        &                 & chain      &                \\
\hline
        & There are       & Spherical  & Elliptic       \\
Comments& two Lax         & bare       & bare           \\
        & representations & curve      & curve          \\
\hline
\end{tabular}
\end{center}
\end{figure}
\vspace{10pt}

\noindent
This table reflects several possible generalizations of the periodic Toda
chain. First of all, one can extent the representation in terms of the "large"
Lax matrices (\ref{Lax}). It naturally leads to the system whose potential is
doubly-periodic function of coordinates. This system is the Calogero-Moser
system. It is associated with the elliptic bare spectral curve and generating
1-form $dS=\lambda d\xi$, $\xi$ being the coordinate on the bare torus. On
physical side, this system corresponds to including the adjoint matter
hypermultiplets \cite{Cal}.

The second possible extension is generalization of the $2\times 2$
Lax representation of the Toda chain (\ref{TodaLax}). This leads to the $XXX$
spin chain with the cylindrical bare spectral curve and the same generating
1-form $dS=\lambda dw/w$. Physically, this system describes fundamental
matter hypermultiplets included \cite{GMMM,GGM1}.

Now, each of these two systems can be further generalized to higher
dimensional, $5d$ and $6d$ SYM gauge theories \cite{hd}-\cite{hd1}, with the
target space (the fifth and sixth dimensions) being accordingly a cylinder
and a torus. Within the integrable framework, this means putting momenta of
the system onto the cylinder (Hamiltonians periodic in momenta) and the torus
(Hamiltonians doubly-periodic in momenta) respectively.  In the generating 1-form one needs just to substitute
respectively $\lambda\to\log\lambda$ in $5d$ and $\lambda\to\zeta$ in $6d$,
$\zeta$ being the coordinate on the target space torus.
At the same time, the
bare spectral curve associated with coordinate dependence is not changed
with this extension.

In a word, adding adjoint matter makes the coordinate dependence (and the
bare spectral curve) elliptic, while coming to higher dimensions provides
trigonometric and elliptic momentum dependencies (and the corresponding target
space).

Let us discuss a little the dressed spectral curves.
In the adjoint matter case, the spectral curve is non-hyperelliptic, since the
bare curve is elliptic. Therefore, it can be described as some covering of
the hyperelliptic curve. We do not go into
further details here, just referring to \cite{WDVVlong,BMMM2}, since the WDVV
equations do not hold, at least, in the standard form (\ref{wdvv}), with the
elliptic bare curve (see below).

Instead, we describe the dressed spectral curves for the $4d$ theories without
adjoint matter with the classical gauge groups
in more explicit terms. Let us note that in all these cases the
curves are hyperelliptic, since all of them follow from some $2\times 2$ Lax
representation and, therefore, are the spectral curves of the form
(\ref{Todasc}).

More concretely, for the SYM gauge theory with the gauge group $G$, one should
consider the integrable system given by this $2\times 2$ Lax representation on
the Dynkin diagram for the corresponding dual affine algebra ${\hat
G}^{\vee}$ \cite{GM}.

The spectral curves can be described by the general formula
\be
{\cal P}(\lambda, w) = 2P(\lambda) - w - \frac{{\cal Q}(\lambda)}{w}
\label{curven}
\ee
Here $P(\lambda)$ is the characteristic polynomial of the
algebra $G$ itself for all $G\ne C_n$, i.e.
\be
P(\lambda) = \det(G - \lambda I) =
\prod_i (\lambda - \lambda_i)
\ee
where determinant is taken in the first fundamental representation
and $\lambda_i$'s are the eigenvalues of the algebraic element $G$.
For the pure gauge theories with the classical groups \cite{MW},
${\cal Q}(\lambda)=\lambda^{2s}$ and\footnote{In the symplectic
case, the curve can be easily recast in the form with polynomial
$P(\lambda)$ and $s=0$.}
\be
A_{n-1}:\ \ \ P(\lambda) = \prod_{i=1}^{n}(\lambda - \lambda_i), \ \ \
\ s=0;\\
B_n:\ \ \ P(\lambda) = \lambda\prod_{i=1}^n(\lambda^2 - \lambda_i^2),
\ \ \ s=2;\\
C_n:\ \ \ P(\lambda) = \prod_{i=1}^n(\lambda^2 - \lambda_i^2)-
{2\over\lambda^2},
\ \ \ s=-2;\\
D_n:\ \ \ P(\lambda) = \prod_{i=1}^n(\lambda^2 - \lambda_i^2),
\ \ \ s=2
\label{charpo}
\ee
For exceptional groups, the curves arising from the characteristic
polynomials of the dual affine algebras do not acquire the hyperelliptic
form although the WDVV equations seem to be still fulfilled.

In order to include $n_F$ massive hypermultiplets in the
first fundamental representation one can just change
$\lambda^{2s}$ for ${\cal Q}(\lambda) = \lambda^{2s}
\prod_{\iota = 1}^{n_F} (\lambda - m_\iota)$ if $G=A_n$
and for ${\cal Q}(\lambda) = \lambda^{2s} \prod_{\iota =
1}^{n_F}(\lambda^2 - m^2_\iota)$ if $G=B_n,C_n,D_n$ \cite{dual,fuma,GM}.

Note that the $5d$ theories can be also described by the same curves
but by different 1-forms $dS$ \cite{hd1}.

\section{WDVV equations in SW theories}
As we already discussed, in order to derive the WDVV equations along the line
used in the context of the LG theories, we need three crucial ingredients:
flat moduli, residue formula and associative algebra. However, the first two
of these are always contained in the SW construction provided the
underlying integrable system is known. Indeed, one can derive (see
\cite{WDVVlong}) the
following residue formula
\be\label{resSW}
F_{ijk}=
\stackreb{d\omega= 0}{\hbox{res}}
\frac{d\omega_id\omega_jd\omega_k}{d\omega d\lambda}
\ee
where the proper flat moduli $a_i$'s are given by formula (\ref{defprep}).
Thus, the only point to be checked is the existence of the associative
algebra. The residue formula (\ref{resSW}) hints that this algebra is to be
the algebra $\Omega^1$ of the holomorphic differentials $d\omega_i$. In the
forthcoming discussion we restrict ourselves to the case of pure gauge
theory, the general case being treated in complete analogy.

Let us consider the algebra $\Omega^1$ and fix three differentials $dQ$,
$d\omega$, $d\lambda\ \in\Omega^1$. The product in this algebra is given
by the expansion
\be\label{prod}
d\omega_id\omega_j=C^k_{ij}d\omega_kdQ+(\ast)d\omega+(\ast)d\lambda
\ee
that should be factorized over the ideal spanned by the differentials
$d\omega$ and $d\lambda$.
This product belongs to the space of quadratic holomorphic differentials:
\be
\Omega^1\cdot\Omega^1\in\Omega^2\cong\Omega^1\cdot\left(dQ
\oplus d\omega\oplus d\lambda\right)
\ee
Since the dimension of the space of quadratic holomorphic differentials is
equal to $3g-3$, the l.h.s. of (\ref{prod}) with arbitrary $d\omega_i$'s is
the vector space of dimension $3g-3$. At the same time, at the
r.h.s. of (\ref{prod}) there are $g$ arbitrary coefficients $C_{ij}^k$ in the
first term (since there are exactly so many holomorphic 1-differentials that
span the arbitrary holomorphic 1-differential $C_{ij}^kd\omega_k$), $g-1$
arbitrary holomorphic differentials in the second term (one differential
should be subtracted to avoid the double counting) and $g-2$ holomorphic
1-differentials in the third one. Thus, totally we get that the r.h.s. of
(\ref{prod}) is spanned also by the basis of dimension $g+(g-1)+(g-2)=3g-3$.

This means that the algebra exists in the general case of the SW construction.
However, generally this algebra is not associative. This is because, unlike
the LG case, when it was the
algebra of polynomials and, therefore, the product
of the two belonged to the same space (of polynomials), product in the algebra
of holomorphic 1-differentials no longer belongs to the same space but to the
space of quadratic holomorphic differentials. Indeed, to check associativity,
one needs to consider the triple product of $\Omega^1$:
\be\label{assSW}
\Omega^1\cdot\Omega^1\cdot\Omega^1\in\Omega^3=
\Omega^1\!\cdot
\left(dQ\right)^2\oplus\Omega^2\!\cdot d\omega\oplus\Omega^2
\!\cdot d\lambda
\ee
Now let us repeat our calculation: the dimension of the l.h.s. of this
expression is $5g-5$ that is the dimension of the space of holomorphic
3-differentials. The dimension of the first space in expansion of the r.h.s.
is $g$, the second one is $3g-4$ and the third one is $2g-4$. Since
$g+(3g-4)+(2g-4)=6g-8$ is greater than $5g-5$ (unless $g\le 3$), formula
(\ref{assSW}) {\bf does not} define the unique expansion of the triple
product of $\Omega^1$ and, therefore, the associativity spoils.

The situation can be improved if one considers the curves with additional
involutions. As an example, let us consider the family of hyperelliptic
curves: $y^2=Pol_{2g+2}(\lambda)$. In this case, there is the involution,
$\sigma:\ y\to -y$ and $\Omega^1$ is spanned by the $\sigma$-odd holomorphic
1-differentials ${x^{i-1}dx\over y}$, $i=1,...,g$. Let us also note that both
$dQ$ and $d\omega$ are $\sigma$-odd, while $d\lambda$ is $\sigma$-even. This
means that $d\lambda$ can be only meromorphic on the surface without
punctures (which is, indeed, the case in the absence of mass
hypermultiplets). Thus,
$d\lambda$ omits from formula (\ref{prod}) that acquires the form
\be\label{prodhe}
\Omega^2_+=\Omega^1_-\cdot dQ\oplus\Omega^1_-\cdot d\omega
\ee
where we expanded the space of
holomorphic 2-differentials into the parts with definite $\sigma$-parity:
$\Omega^2=\Omega^2_+\oplus\Omega^2_-$, which are manifestly given by the
differentials ${x^{i-1}(dx)^2\over y^2}$, $i=1,...,2g-1$ and
${x^{i-1}(dx)^2\over y}$, $i=1,...,g-2$ respectively.  Now it is easy to
understand that the dimensions of the l.h.s. and r.h.s. of (\ref{prodhe})
coincide and are equal to $2g-1$.

Analogously, in this case, one can check the associativity. It is given by the
expansion
\be
\Omega^3_-=\Omega_-^1\cdot \left(dQ\right)^2\oplus\Omega_+^2\cdot d\omega
\ee
where both the l.h.s. and r.h.s. have the same dimension: $3g-2=g+(2g-2)$.
Thus, the algebra of holomorphic 1-differentials on hyperelliptic curve
is really associative. This completes the proof of the WDVV equations in this
case.

Now let us briefly look at cases when there exist the associative algebras
basing on the spectral curves discussed in the previous section. First of
all, it exists in the theories with the gauge group $A_n$, both in the pure
gauge $4d$ and $5d$ theories and in the theories with fundamental matter,
since, in accordance with the previous section, the corresponding spectral
curves are hyperelliptic ones of genus $n$.

The theories with the gauge groups $SO(n)$ or $Sp(n)$ are also described
by the hyperelliptic curves. The curves, however, are of higher genus
$2n-1$. This would naively destroy all the reasoning of this section.
The arguments,
however, can be restored by noting that the corresponding curves (see
(\ref{charpo})) have yet {\bf another} involution, $\rho:\
\lambda\to-\lambda$. This allows one to expand further the space of
holomorphic differentials into the pieces with definite $\rho$-parity:
$\Omega^1_-=\Omega^1_{--}\oplus\Omega^1_{-+}$ etc. so that the proper algebra
is generated by the differentials from $\Omega^1_{--}$. One can easily check
that it leads again to the associative algebra.

Consideration is even more tricky for the exceptional groups, when the
corresponding curves are looking non-hyperelliptic. However, additional
symmetries should allow one to get associative algebras in these cases too.

There are more cases when the associative algebra exists. First of all, these
are $5d$ theories, with and without fundamental matter \cite{WDVVlong}. One
can also consider the SYM theories with gauge groups being the product of
several factors, with matter in the bi-fundamental representation \cite{W}.
These theories are described by $SL(p)$ spin chains \cite{GGM1} and the
existence of the associative algebra in this case has been checked in
\cite{Isid}.

The situation is completely different in the adjoint matter case. In four
dimensions, the theory is described by the Calogero-Moser integrable
system. Since, in this case, the curve is non-hyperelliptic and has no
enough symmetries, one needs to include into consideration both the
differentials $d\omega$ and $d\lambda$ for algebra to exist. However, under
these circumstances, the algebra is no longer to be associative how it was
demonstrated above. This can be done also by direct calculation for several
first values of $n$ (see \cite{WDVVlong}). This also explains the lack of the
perturbative WDVV equations in this case (see sect.2).

\section{Covariance of the WDVV equations}
After we have discussed the role of the (generalized) WDVV equations in SYM
gauge theories of the Seiberg-Witten type, let us briefly describe the
general structure of the equations themselves. We look at
them now just as at some over-defined set of non-linear equations for a
function (prepotential) of $r$ variables\footnote{We deliberately
choose different notations for these variables, $t$ instead of $a$
in the gauge theories,
in order to point out more general status
of the discussion.} (times),
$F(t^i)$, $i=1,\ldots,r$, which can
 be written in the form (\ref{WDVVgen})
\be
F_i G^{-1} F_j = F_j G^{-1} F_i, \nn \\
G = \sum_{k=1}^r \eta^k F_k, \ \ \
\forall i,j = 1,\ldots,r \ \ {\rm and} \ \
\forall \eta^k(t)
\label{WDVV}
\ee
$F_i$ being $r\times r$ matrices
$\displaystyle{(F_i)_{jk} = F_{,ijk} = \frac{\partial^3 F}
{\partial t^i\partial t^j\partial t^k}}$
and the "metric" matrix $G$ is an arbitrary linear
combination of $F_k$'s, with coefficients
$\eta^k(t)$ that can be time-dependent.\footnote{We already discussed
in sect.2 that one can add to the set of times (moduli) in the WDVV
equations the parameter $\Lambda$ \cite{BM}. In this case,
the prepotential that depends on one extra variable $t^0\equiv\Lambda$
can be naturally considered as a homogeneous function of degree $2$:
$$
{\cal F}(t^0,t^1,\ldots,t^r) = (t^0)^2F(t^i/t^0),
$$
see \cite{whit} for the general theory.
As explained in \cite{BM}, the WDVV equations
(\ref{WDVV}) for $F(t^i)$ can be also rewritten
in terms of ${\cal F}(t^I)$:
$$
{\cal F}_I \hat{\cal G}^{-1} {\cal F}_J =
{\cal F}_J \hat{\cal G}^{-1} {\cal F}_I,  \ \ \
\forall I,J = 0,1,\ldots,r; \{ \eta^K(t)\}
$$
where this time ${\cal F}_I$ are $(r+1)\times(r+1)$ matrices
of the third derivatives of ${\cal F}$ and
$$
{\cal G} = \sum_{k=0}^r \eta^K {\cal F}_K, \ \ \
\hat{\cal G}^{-1} = (\det {\cal G}) {\cal G}^{-1}
$$
Note that the homogeneity of ${\cal F}$ implies that
$t^0$-derivatives are expressed through those w.r.t. $t^i$, e.g.
$$
t^0{\cal F}_{,0ij}=-{\cal F}_{,ijk}t^k,\ \ \ \
t^0{\cal F}_{,00i}={\cal F}_{,ikl}t^kt^l,\ \ \ \
t^0{\cal F}_{,000}=-{\cal F}_{,klm}t^kt^lt^m \ \ \ \hbox{etc.}
$$
Thus, all the "metrics" ${\cal G}$ are degenerate, but
$\hat{\cal G}^{-1}$ are non-degenerate.
One can easily reformulate the entire present section in terms of
${\cal F}$.Then, e.g., the Baker-Akhiezer vector-function
$\psi(t)$ should be just substituted by the manifestly
homogeneous (of degree 0) function $\psi(t^i/t^0)$.
The extra variable $t^0$ should not be mixed with the
distinguished "zero-time" associated with the constant metric
in the $2d$ topological theories which generically
does not exist (when it does, see comment 2.3 below,
we identify it with $t^r$).
\label{f1}
}

The WDVV equations imply consistency of the following system
of differential equations \cite{WDVVv1}:
\be
\left( F_{,ijk}\frac{\partial}{\partial t^l} -
       F_{,ijl}\frac{\partial}{\partial t^k} \right)\psi^j(t) = 0,
\ \ \ \forall i, j, k
\label{ls}
\ee
Contracting with the vector $\eta^l(t)$, one can also rewrite
it as
\be\label{*}
\frac{\partial \psi^i}{\partial t^k} = C^i_{jk} D\psi^j,
\ \ \ \forall i, j
\ee
where
\be\label{3}
C_k = G^{-1}F_k, \ \ \ G = \eta^lF_l,\ \ \ D = \eta^l\partial_l
\ee
(note that the matrices $C_k$ and the differential $D$ depend on
choice of $\{\eta^l(t)\}$, i.e. on choice of the metric $G$)
and (\ref{WDVV}) can be rewritten as
\be\label{***}
\left[C_i,C_j\right]=0,
\ \ \ \forall i, j
\ee

As we already discussed,
the set of the WDVV equations (\ref{WDVV}) is invariant under
{\it linear} change of the time variables with the prepotential
unchanged \cite{WDVVlong}.
According to the second paper of \cite{typeB} and especially to \cite{L},
there can exist also {\it non-linear} transformations
which preserve the WDVV structure, but they generically change
the prepotential.
In \cite{WDVVv2}, it is shown that such transformations are naturally
induced by solutions of the linear system (\ref{ls}):
\be
t^i \ \longrightarrow \ \tilde t^i = \psi^i(t), \nn \\
F(t) \ \longrightarrow \ \tilde F(\tilde t),
\label{tr}
\ee
so that the period matrix remains intact:
\be
F_{,ij} = \frac{\partial^2 F}{\partial t^i\partial t^j}
= \frac{\partial^2 \tilde F}{\partial \tilde t^i\partial \tilde t^j}
\equiv \tilde F_{,\hat i\hat j}
\label{pm}
\ee

Now let us make some comments.

\noindent
{\bf 1.} As explained in \cite{WDVVv1}, the linear system (\ref{ls})
has infinitely many solutions. The "original" time-variables
are among them: $\psi^i(t) = t^i$.

\noindent
{\bf 2.} Condition (\ref{pm}) guarantees that the transformation
(\ref{tr}) changes the linear system (\ref{ls}) only by
a (matrix) multiplicative factor, i.e. the set of solutions
$\{\psi^i(t)\}$ is invariant of (\ref{tr}). Among other
things this implies that successively applying (\ref{tr})
one does not produce new sets of time-variables.

\noindent
{\bf 3.} We already discussed that, in the case of $2d$ topological
models \cite{typeA,typeB,L}, there is a distinguished time-variable,
say, $t^r$, such that all $F_{,ijk}$ are independent of $t^r$:
\be
\frac{\partial}{\partial t^r} F_{ijk} = 0 \ \ \
\forall i,j,k = 1,\ldots, r
\ee
(equivalently, $\frac{\partial}{\partial t^i} F_{rjk} = 0$
$\forall i,j,k$).
Then, one can make the Fourier transform of (\ref{ls})
with respect to $t^r$ and substitute it by the system
\be
\frac{\partial}{\partial t^j} \hat\psi^i_z = zC^i_{jk}
\hat\psi^k_z,
\ \ \ \forall i, j
\ee
where
$\hat\psi^k_z(t^1,\ldots,t^{r-1}) =
\int \psi^k_z(t^1,\ldots,t^{r-1},t^r) e^{zt^r}dt^r$.
In this case, the set of transformations (\ref{tr})
can be substituted by a family, labeled by a single variable $z$:
\be
t^i \ \longrightarrow \ \tilde t^i_z = \hat\psi^i_z(t)
\ee
In the limit $z \rightarrow 0$ and for the particular choice of the
metric, $\check G=F_r$, one obtains the particular
transformation
\be
{\partial \tilde t^i \over \partial t^j}
={\check C}^i_{jk}h^k, \ \ \ h^k = const,
\label{L}
\ee
discovered in \cite{L}. (Since ${\check C}_i=\partial_j \check C$, one can
also write ${\check C}_i={\check C}^i_kh^k$, ${\check
C}^i_k=\left(F_r^{-1}\right)^{il}F_{,lk}$.)

\noindent
{\bf 4.} Parameterization like (\ref{L}) can be used in
the generic situation
(\ref{tr}) as well (i.e. without distinguished $t^r$-variable and for
the whole family (\ref{tr})), the only change is that
$h^k$ is no longer a constant,
but a solution to
\be\label{10}
\left(\partial_j - DC_j\right)^i_kh^k = 0
\ee
($h^k = D\psi^k$ is always a solution, provided $\psi^k$
satisfies (\ref{*})).

Note also that, although we have described a set of non-trivial non-linear
transformations which preserve the structure of the WDVV
equations (\ref{WDVV}), the consideration above does not
{\it prove} that {\it all} such transformations are of the
form (\ref{tr}), (\ref{pm}).
Still, (\ref{tr}) is already unexpectedly large,
because (\ref{WDVV}) is an {\it over-defined} system and
it could seem to be very {\it restrictive}, if to have any solutions
at all.

\section{Concluding remarks}
To conclude this short review, let us emphasize that a lot of
problems have to be solved before we get any real understanding of what the
WDVV structure means. We already mentioned the problem of lack of the WDVV
equations for the Calogero-Moser system.
The way to resolve this problem might be to construct higher
associativity conditions like it has been done by E.Getzler in the elliptic
case \cite{Getzler}, that is to say, for the elliptic Gromov-Witten
clas\-ses.
The other kind of problem is that the WDVV equations in the
type A topological theories themselves do still wait for the explanation in
terms of associative algebras.

All these problems are to be resolved in order
to establish to what extent there is a really deep reason for the WDVV
equations to emerge in topological and Seiberg-Witten theories.
Note that the latter two are connected through the Whitham hierarchies
(see, e.g., \cite{Don}, the second paper in \cite{whit} and the review by
A.Morozov at the Workshop) and there are also tight connections of the
Whitham hierarchies with the WDVV equations \cite{typeB}.

The other problem is more on the structure of the WDVV equations themselves:
up to now, we do not understand what is the class of solutions to the
equations and how wide it is, having some particular examples in
hands. And maybe even more obscure are the {\it origins and implications} of
the covariance of the WDVV equations.

At last, we still have no any clear
understanding of connections between the WDVV and integrable structures. What
we know is rather a set of random observations.

All these problems appeal for better understanding before we could put the
associative algebras and WDVV equations on any solid footing.

\vspace{1cm}

I am grateful to H.W.Braden, A.Gorsky, A.Marshakov and A.Moro\-zov
for useful discussions and to H.W.Braden for strong encouraging me to write
this review. I also acknowledge the hospitality of University of Edinburgh
and the Royal Society for support under a joint project.

The research is partly supported by the
RFBR grant 98-01-00328,
INTAS grant 96-482 and
the program for support of the scientific schools 96-15-96798.


\begin{thebibliography}{12}

\bibitem{SW} N.Seiberg and E.Witten, Nucl.Phys., {\bf B426} (1994) 19,
hep-th/9407087; Nucl.Phys., {\bf B431} (1994) 484,
hep-th/9408099

\bibitem{GKMMM} A.Gorsky, I.Krichever, A.Marshakov, A.Mironov and A.Morozov,
Phys.Lett., {\bf B355} (1995) 466, hep-th/9505035

\bibitem{MW}
E.Martinec and N.Warner,
Nucl.Phys., {\bf B459} (1996) 97-112, hep-th/9509161

\bibitem{GMMM} A.Gorsky, A.Marshakov, A.Mironov and A.Morozov,
Phys.Lett., {\bf B380} (1996) 75, hep-th/9603140

\bibitem{GGM1}
A.Gorsky, S.Gukov and A.Mironov, Nucl.Phys., {\bf B517} (1998)
409, hep-th/9707120

\bibitem{int}
T.Nakatsu and K.Takasaki, Mod.Phys.Lett., {\bf 11} (1996) 417,
hep-th/9509162;\\
T.Eguchi and S.K.Yang, Mod.Phys.Lett.,
{\bf A11} (1996) 131-138, hep-th/9510183;\\
E.Martinec and N.Warner, hep-th/9511052;\\
A.Marshakov, Mod.Phys.Lett., {\bf A11} (1996) 1169, hep-th/9602005;\\
C.Ahn and S.Nam, Phys.Lett., {\bf B387} (1996) 304, hep-th/9603028;\\
K.Takasaki, solv-int/9704004; solv-int/9705016;\\
I.Krichever and D.Phong, hep-th/9708170

\bibitem{GM} A.Gorsky and A.Mironov, hep-th/9902030; to appear in Nuclear
Physics, B

\bibitem{Cal}
R.Donagi and E.Witten, Nucl.Phys., {\bf B460} (1996) 299,
hep-th/9510101;\\
E.Martinec, Phys.Lett., {\bf B367} (1996) 91, hep-th/9510204;\\
A.Gorsky and A.Marshakov, Phys.Lett., {\bf B375} (1996) 127,
hep-th/9510224;\\
H.Itoyama and A.Morozov, Nucl.Phys., {\bf B477} (1996) 855, hep-th/9511125;
hep-th/9512161

\bibitem{N} N.Nekrasov, Nucl.Phys., {\bf B531} (1998) 323, hep-th/9609219

\bibitem{hd} A.Gorsky, A.Marshakov, A.Mironov and A.Morozov,
hep-th/9604078;\\
A.Gorsky, S.Gukov and A.Mironov, Nucl.Phys., {\bf B518}
(1998) 689, hep-th/9710239

\bibitem{BMMM2}
H.W.Braden, A.Marshakov, A.Mironov and A.Morozov, hep-th/9902205

\bibitem{hd1} A.Marshakov and A.Mironov, Nucl.Phys., {\bf B518}
(1998) 59-91, hep-th/9711156

\bibitem{rev} For reviews see: \\
H.Itoyama and A.Morozov, hep-th/9601168;\\
A.Marshakov, Int.J.Mod.Phys., {\bf A12} (1997), 1607,
hep-th/9610242; Theor.\& Math.Phys., {\bf 112} (1997) 791, hep-th/9702083;\\
A.Mironov, hep-th/9801149;\\
R.Donagi, alg-geom/9705010;\\
A.Klemm, hep-th/9705131;\\
S.Ketov, hep-th/9710085;\\
C.Gomez and R.Hernandez, hep-th/9711102;\\
R.Carroll, hep-th/9712110

\bibitem{WDVVa}
A.Marshakov, A.Mironov and A.Morozov, Phys.Lett., {\bf B389} (1996)
43, hep-th/9607109

\bibitem{WDVVlong} A.Marshakov, A.Mironov and A.Morozov,
hep-th/9701123

\bibitem{WDVVr}
A.Marshakov, A.Mironov and A.Morozov,
Mod.Phys.Lett., {\bf A12} (1997) 773, hep-th/9701014;\\
A.Mironov, hep-th/9704205

\bibitem{WDVVv1}
A.Morozov, Phys.Lett., {\bf B427} (1998) 93-96, hep-th/9711194

\bibitem{WDVVv2}
A.Mironov and A.Morozov, Phys.Lett., {\bf B424} (1998) 48-52,
hep-th/9712177

\bibitem{BMMM1} H.W. Braden, A.Marshakov, A.Mironov and A.Morozov,
hep-th/9812078; to appear in Phys.Lett., B

\bibitem{WDVV}
E.Witten, Surv.Diff.Geom. {\bf 1} (1991) 243;\\
R.Dijkgraaf, E.Verlinde, H.Verlinde, Nucl.Phys.,
{\bf B352} (1991) 59

\bibitem{typeA}
Yu.Manin, {\sl Frobenius manifolds, quantum cohomology and moduli spaces},
MPI, 1996;\\
M.Kontsevich, Yu.Manin, Comm.Math.Phys., {\bf 164} (1994) 525

\bibitem{typeB}
I.Krichever, Comm.Pure Appl.Math., {\bf 47} (1994) 437;\\
B.Dubrovin,  Nucl.Phys., {\bf B379} (1992) 627; hep-th/9407018

\bibitem{whit}
I.Krichever, Comm.Math.Phys., {\bf 143}
(1992) 415, hep-th/9205110;\\
A.Gorsky, A.Marshakov, A.Mironov, A.Morozov, Nucl.Phys., {\bf B527}
(1998) 690-716, hep-th/9802007

\bibitem{IY} K.Ito and S.-K.Yang, Phys.Lett., {\bf B433} (1998) 56-62

\bibitem{Ves} A.Veselov, hep-th/9902142

\bibitem{dual} P.Argyres and A.Shapere, Nucl.Phys.,
{\bf B461} (1996) 437, hep-th/9509175

\bibitem{Sei} N.Seiberg, Phys.Lett., {\bf B388} 753; hep-th/9608111;\\
K.Intriligator, D.R.Morrison and N.Seiberg,
Nucl.Phys., {\bf B497} (1997) 56; hep-th/9702198;\\
N.Seiberg and D.Morrison, Nucl.Phys., {\bf B483} (1997) 229;
hep-th/9609070

\bibitem{HM} J.Harvey and G.Moore, Nucl.Phys., {\bf B463} (1996)
315-368; hep-th/9510182

\bibitem{BM} G.Bertoldi and M.Matone, Phys.Rev., {\bf D57}
(1998) 6483-6485; hep-th/9712109

\bibitem{fuma}
A.Hanany, Nucl.Phys., {\bf 466} 85, hep-th/9509176;\\
U.Danielsson and B.Sundborg, Phys.Lett., {\bf B358} (1995) 273,
hep-th/9504102;\\
A.Brandhuber and K.Landsteiner, Phys.Lett., {\bf B358} (1995) 73,
hep-th/9507008

\bibitem{W} E.Witten, Nucl.Phys., {\bf B500} (1997) 3, hep-th/9703166

\bibitem{Isid} J.M.Isidro, Nucl.Phys., {\bf B539} (1999) 379-402,
hep-th/9805051

\bibitem{L} A.Losev, JETP Lett., {\bf 65} (1997) 374.

\bibitem{Getzler}
L.Caporaso, J.Harris, alg-geom/9608025;\\
E.Getzler, alg-geom/9612004;\\
for the latest review and further references, see:\\
E.Getzler, math/9812026

\bibitem{Don} G.Moore and E.Witten, hep-th/9709193;\\
A.Losev, N.Nekrasov and S.Shatashvili, Nucl.Phys., {\bf B534} (1998) 549-611,
hep-th/9711108; hep-th/9801061;\\
G.Moore and M.Mari\~no, Nucl.Phys.Procl.Suppl., {\bf 68} (1998)
336-347, hep-th/9712062

\end{thebibliography}
\end{document}